# Charge distribution at metal-multilayered semiconductor interfaces

Qian Wang,[1,2] Yangfan Shao,[1,3] and Xingqiang Shi[1,a]

[1] Department of Physics and Guangdong Provincial Key Laboratory for Computational Science and Material Design, Southern University of Science and Technology, Shenzhen 518055, China

[2] Harbin Institute of Technology, Harbin 150080, China

[3] Joint Key Laboratory of the Ministry of Education, Institute of Applied Physics and Materials Engineering, University of Macau, Macau, China

[a] E-mail: shixq@sustech.edu.cn

## Abstract

Thicknesses-dependent performances of metal–multilayered semiconductor junctions have attracted increasing attention, but till present, the mechanism of interaction and the resulting charge distribution at interfaces which control the Schottky barrier and band offset between the semiconductor layers have not been systematically studied. Based on first-principles calculations, the nature and strength of the non-bonding interactions at Metal-$MoS_2$ (M-S) and $MoS_2$-$MoS_2$ (S-S) interfaces in metal-multilayered $MoS_2$ are investigated. We show that the charge distribution at M-S interfaces depends sensitively on the dimensionality and work function of metal substrates: 1) push back effect and metal induced gap states play a main role at 3D metal-$MoS_2$ interfaces; 2) charge transfer occurs in $Mo_2C(OH)_2$ (or $Mo_2CO_2$)-$MoS_2$ interfaces which means electron distribution is determined by the band alignment of metal and $MoS_2$; 3) covalent-like feature appears at $Mo_2CF_2$-$MoS_2$ interface. The S-S interface inherit the charge redistribution at M-S interface for 2D metal-2L $MoS_2$ junction, and have a depinning effect for M-S interface in 3D metal-2L $MoS_2$ junction. We are trying to start drawing general conclusions and developing new concepts to understand metal–multilayered semiconductor interfaces in the strong interaction limit, where charge-transfer effects must be taken into consideration in this paper.

# Introduction

Transition-metal dichalcogenides (TMDs) have been investigated in recent years as a potential semiconductor technology competitor for applications such as electronics [1-5] and optoelectronics [6-8]. $MoS_2$ is one of the representatives which shows strong optical absorption [9,10], and high on/off current ratio[11]. Monolayer $MoS_2$ has a direct band gap (∼1.8–1.9 eV), when van der Waals $MoS_2$ layer added, the layer number dependent bandgap turned from direct to indirect with band gap size decreases at the same time.

However, when contacting 2D $MoS_2$ to metal electrodes, unexpectedly high interface resistance is produced, which deiced the charge carrier transport behaviors across the junction and then significantly impact device performance [12]. The most important parameter for the interface resistance is the Schottky barrier height (SBH), which ideally depends on the energy level alignment of $MoS_2$ and metal for Schottky-Mott limit [13]. Realistically, there is a serious deviation between SBH and Schottky-Mott limit especially for $MoS_2$ absorbed on 3D metal electrodes which is because of the effect of strong Fermi level pinning (FLP) caused by metal induced gap states and the charge redistribution at interface even for a high-quality interface [14]. It's worth noting that 2D metal electrodes can effectively regulate SBH since the 2D metal weakly bonded to 2D semiconductors through van der Waals (vdW) force since the absence of dangling bonds[14,15]. Although there has been a lot of research about metal-$MoS_2$ heterojunctions, a detailed interface study of 2D $MoS_2$ absorbed on 3D and 2D metal substrates is not yet available. What's more, although thicknesses-dependent performance in metal–multilayered semiconductor junctions have attracted increasing attention recently[16-18], clear explanations about semiconductor interlayer interaction have not been reported till present.

Here, via first-principles calculations, we focus on Metal-$MoS_2$ (M-S) interface in Metal-1L $MoS_2$ and $MoS_2$-$MoS_2$ (S-S) interface in Metal-2L $MoS_2$, including 3D metal (e,g, Pt, Ag) and 2D metallic MXenes electrodes. In particular, for frequently-used 3D metal electrodes, a sizable n-type Schottky barriers (SB) often formed accompanied with a strong binding between metal and $MoS_2$ [19,20]. With modulated metal work function, 2D metal transition metal carbides (MXenes) contact to $MoS_2$ with vanishing SB (both n-type and p-type)[21,15]. The SB is closely related to the interface redistribution of charge due to interfacial interactions and change with it accordingly. We have considered three types 2D MXenes $Mo_2C(OH)_2$, $Mo_2CF_2$, $MO_2CO_2$ whose work function located at three different level ($W_M$<EA, EA<$W_M$<IE, IE<$W_M$, where $W_M$ is metal work function, EA is $MoS_2$ electron affinity, and IE is $MoS_2$ ionization energy), when adsorb $MoS_2$ on them, different charge redistributions are found due to different mechanisms. Simultaneously, heterojunctions with 3D metal substrates have been studied, push back effect and metal induced gap states (MIGS) dominates the interaction at the M-S interface. Finally, we discuss the interfacial interaction between semiconductor layers and found that the charge distribution at S-S interface is closely related to the work function and dimension of substrates. The charge distribution at S-S interface determined the band alignment of the second layer $MoS_2$ relative to the first one which will further generate a type-II band alignment between these two $MoS_2$ layers. Therefore, a comprehensive theory of interfacial interaction of metal-semiconductor and semiconductor-semiconductor is concluded based on metal-multilayer $MoS_2$.

## Methods

First-principles calculations are carried out using DFT as implemented in the Vienna Ab initio Simulation Package (VASP) with periodic boundary conditions [22-24]. The projected augmented wave (PAW) [25] was adopted to describe the ion–electron interaction. All the calculations were carried out using SCAN+rVV10 (the SCAN meta-GGA density functional with the rVV10 vdW correction)[26,27] which offered a good performance for layered materials compared with experimental results [28]. The cut-off kinetic energy for plane waves was set to 500 eV. The atomic positions were fully relaxed until the force on each atom was less than 0.01 eV $Å^{-1}$, and the energy convergence value between two consecutive steps was chosen as $10^{-4}$ eV. The Brillouin zone (BZ) was sampled in the Monkhorst-Pack scheme with a k-point grid spacing of 0.01 $Å^{-1}$. A vacuum region of 15 Å is added to minimize the interaction between adjacent slabs and a dipole correction is applied to avoid spurious interactions between periodic images of the slab [29].

## Results and discussion

In Schottky-Mott limit, the SBH $\Phi_B$ is obtained by band alignment of the non-interacting subsystems:

$$\Phi_B^e = \Phi_M - EA \ ; \quad \Phi_B^h = IE - \Phi_M \qquad (1),$$

where $\Phi_B^e$ and $\Phi_B^h$ are the SBH for electrons (n-type) and holes (p-type), $\Phi_M$ is the work function of the metal, $EA$ and $IE$ are the electron affinity and ionization potential of the semiconductor, respectively. $IE - EA$ =band gap. These quantities are the intrinsic properties of isolated materials before they form the junction. In this case, $\Phi_B$ is linearly dependent on the $\Phi_M$ of metals and change in the same amounts as the $\Phi_M$ does. In many cases, however, even for weakly interacting systems, the Schottky-Mott limit is not obeyed; the experimentally determined hole and electron injection barriers are different from those calculated using Formula 1. The origin of these differences can be attributed to the existence of an additional interface dipole Δ, which finally improves the SBH $\Phi_B$ to:

$$\Phi_B^e = \Phi_M - EA - \Delta; \quad \Phi_B^h = IE - \Phi_M + \Delta \qquad (2).$$

The interface dipole $\Delta V$ is caused by the charge redistribution at interface, which shown as the electron density at interface. The electron density difference ($\Delta\rho$) is an effective tool to visualizing bonding at metal-MoS$_2$ interface which is defined as the difference of the electron density of the composite full system and the isolated subsystems:

$$\Delta\rho = \rho_{tot} - \rho_M - \rho_{MoS2} \qquad (3),$$

where $\rho_{tot}$, $\rho_M$, and $\rho_{MoS2}$ are electron density of metal-MoS$_2$, metal and free-standing MoS$_2$, respectively. The $\Delta\rho$ is localized around the metal-MoS$_2$ interface, which includes the contributions from the energy level shift induced by charge transfer and the energy level broadening induced by metal, and can further effects the band alignment. The $\Delta\rho$ and electrostatic potential $\Delta V$ (equal to interface dipole Δ) satisfy the Poisson equation, by solving the Poisson equation, a potential step across the interface is achieved:

$$\Delta V = \frac{e^2}{\epsilon_0 A} \iiint z\Delta\rho dxdydz \qquad (4),$$

where $z$ is the distance from the electrode surface, A is the interface area, $\varepsilon_0$ is the dielectric constant of MoS$_2$.

In DFT calculation, $\Delta V$ in this formula is defined as difference between the asymptotic values of the potential left and right of the interface, a simple alternative expression is illustrated as:

$$\Delta V = W_M - W_{M/S} \quad (5),$$

where $W_M$, $W_{M/S}$ are the work functions of the clean metal surface, and of the metal surface covered by $MoS_2$, respectively (see Figure S1).

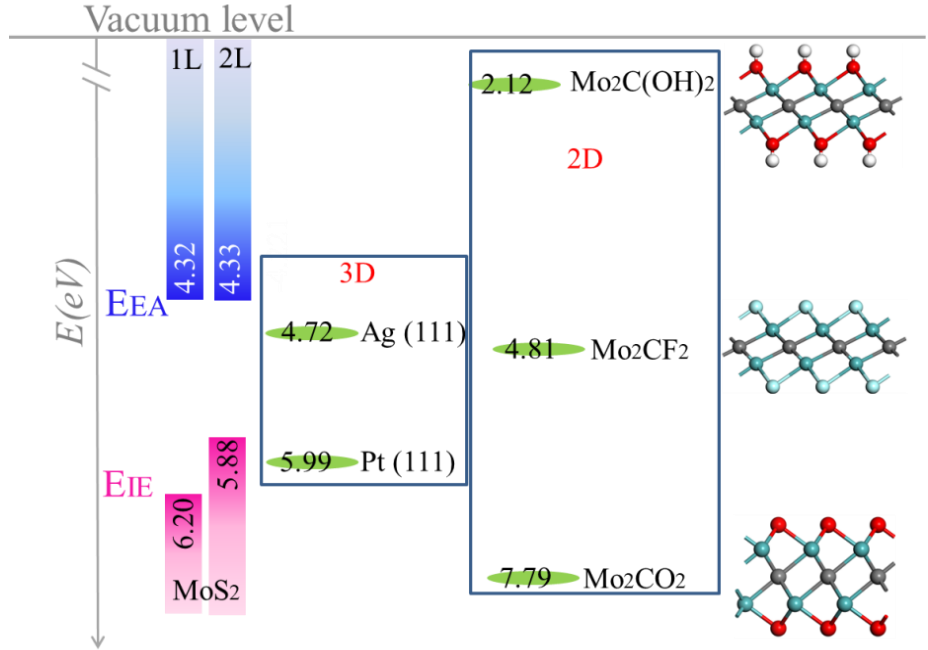

**FIG. 1. Structures and band alignments.** Band alignments of monolayer and bilayer $MoS_2$ and work functions ($W_M$) of 3D and 2D metals; $E_{EA}$ and $E_{IE}$ represent electron affinity and ionization energy of 2D $MoS_2$.

The optimized lattice constant of monolayer (ML) $MoS_2$ is 3.16 Å from the SCAN + rvv10 function, which agrees with theoretical results in literatures [26]. The calculated in-plane lattice constants of metals and the supercell matched metal-$MoS_2$ are shown in Table S. The EA and IE of monolayer and bilayer $MoS_2$ are shown in Figure 1, the EA is 4.32 eV and almost unchanged while IE decrease obviously (from 6.20 eV to 5.88 eV) when the layer number increased from 1 to 2. We align metal into two lines, metals in right line belong to 2D MXenes $Mo_2CT_2$ (T=OH, F, O) whose geometry structures are shown correspondingly. The work function of MXenes is determined by functional groups which produced the effect of surface dipoles [30]: O termination has the largest work function (7.79 eV), OH has the smallest one (2.12 eV), whereas F exhibits a work function of 4.81 eV which locate between energy of EA and IE of $MoS_2$.

In the junction formed between these metals and $MoS_2$, the equilibrium interface distances (D),

defined as the average out-of-plane distance between the S atoms at the bottom of MoS$_2$ and atoms at the surface of metals, are belong to distances of vdW interactions for 2D MXenes, and stronger than pure vdW fore for 3D metal, which is further evidenced binding energies. Here, the binding energy (E$_b$) between metals and the 1L MoS$_2$ is defined as $E_b = (E_{tot} - E_M - E_{MoS2})/A$, where $E_{tot}$, $E_M$, and $E_{MoS2}$ are the total energies of metal-MoS$_2$, metal and free-standing MoS$_2$, respectively; A is the interface area. The calculated E$_b$ values are in the range of -0.26 to -0.82 J/m$^2$ (Table 1) for 2D substrates, which suggests that 1L MoS$_2$ is physisorbed on 2D metals, while -0.62 and -0.96 J/m$^2$ (Table 1) for Ag and Pt.

Table I. Equilibrium Interface Distance (D) and Binding Energy (E$_b$) of 2D Metal−1L MoS$_2$ Junctions.

|  | D (Å) | Absorption energy (J/m$^2$) |
|---|---|---|
| Mo$_2$C(OH)$_2$ | 2.00 | -0.82 |
| Mo$_2$CO$_2$ | 2.82 | -0.37 |
| Mo$_2$CF$_2$ | 3.12 | -0.26 |
| Pt | 2.51 | -0.96 |
| Ag | 2.79 | -0.62 |

It is noticeable that, the strong FLP at the interface of traditional 3D elemental metals (etc. Ag and Pt) and 2D MoS$_2$ becomes very weak at 2D MXenes-MoS$_2$ interface since the weak vdW interaction between them. The vdW interaction at 2D MXenes-MoS$_2$ interface makes the charge distribution between them simple and clear. For 3D metal substrates, the interaction at M-S interface is very complicated, multiple mechanisms work together to determine the charge distribution at the interface.

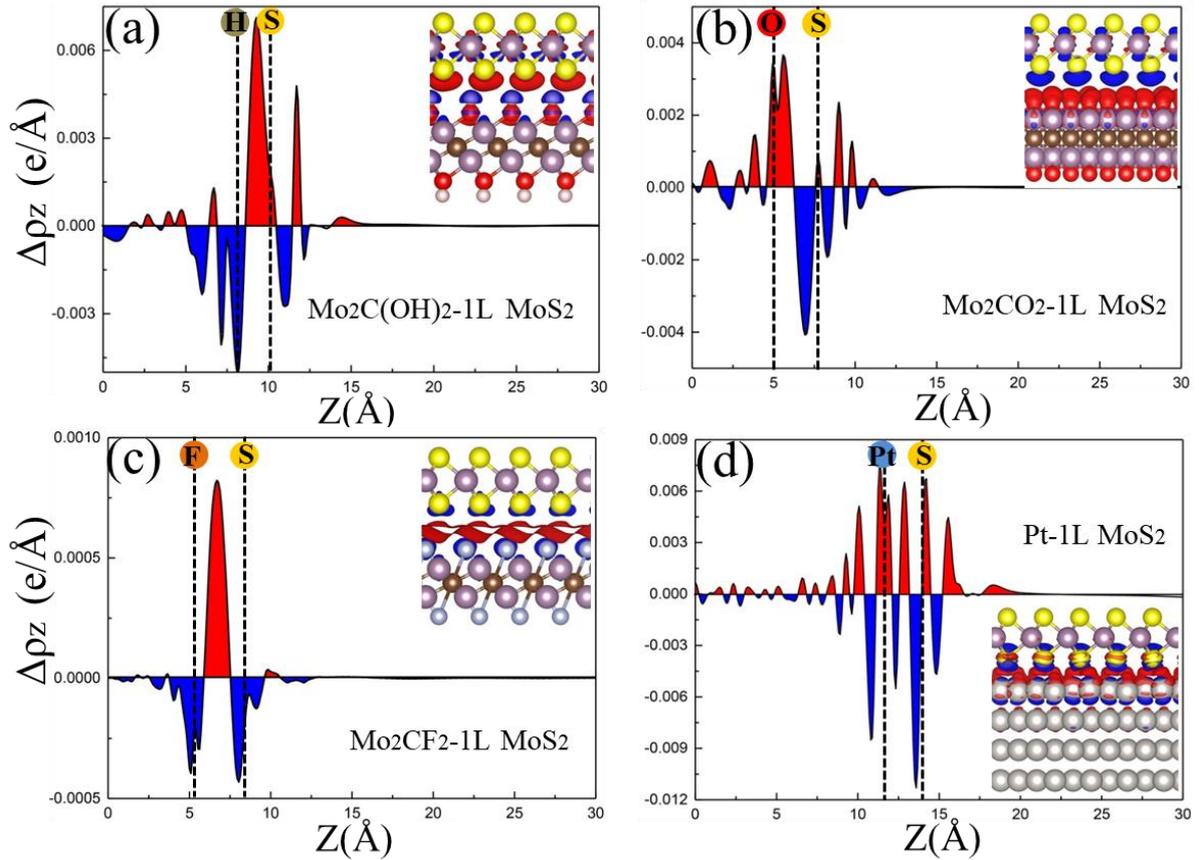

**FIG. 2. Charge distribution at Metal-MoS$_2$ interface.** Plane-averaged electron density difference along the vertical z-direction to the Metal-MoS$_2$ interfaces. Red (blue) regions represent electron accumulation (depletion) regions. The interface between metal and MoS$_2$ is demonstrated within two black dotted lines. The corresponding differential charge densities are shown in the insert images. To ensure that each image is clearly, different isovalue are used for different systems, the isovalue is $10^{-4}$, $3*10^{-3}$, $10^{-3}$, $7*10^{-4}$ |e|/Bohr$^3$ for (a-d), respectively.

The plane-averaged electron density differences along the vertical z-direction to the Metal-MoS$_2$ interfaces are shown in Figure 2. When MoS$_2$ is adsorbed on a metallic surface, the factors that can modify the interface dipole are concluded.

As shown in Figure 2d, for 3D metal-1L MoS$_2$ junction, electrons aggregate at Pt side and deplete at MoS$_2$ side on the whole. But in fact, the charge distribution at the interface is very complicated, which can be seen from the charge undulating at the interface in the Figure 2d. The push back effect plays a main role at 3D metal Pt-1L MoS$_2$ interface, which is generally observed in the physisorption of 2D layers on metal substrates [19,31]. The main reason for this effect is the antisymmetrization of the metal and adsorbate wave functions. When 2D monolayer MoS$_2$ adsorbs onto the metal, the two wave functions overlap. In order to lower the density in the overlap region, a rearrangement of the electron density occurs consequently based on Pauli repulsion. Since the wave function of 3D metal Pt is more extended and deformable than monolayer MoS$_2$, the electrons are pushed back into the metal, leading to the phenomenon that electrons accumulate at Pt side and deplete at MoS$_2$ side. As a result, the work function is effectively lowered when MoS$_2$ absorbed on Pt (111) surface as illustrate in Figure 3d. Such interfacial charge distributions also occur in Ag-MoS$_2$ (Figure S4), and the work function of 3D metal Ag is 4.72 between EA and IE of monolayer MoS$_2$.

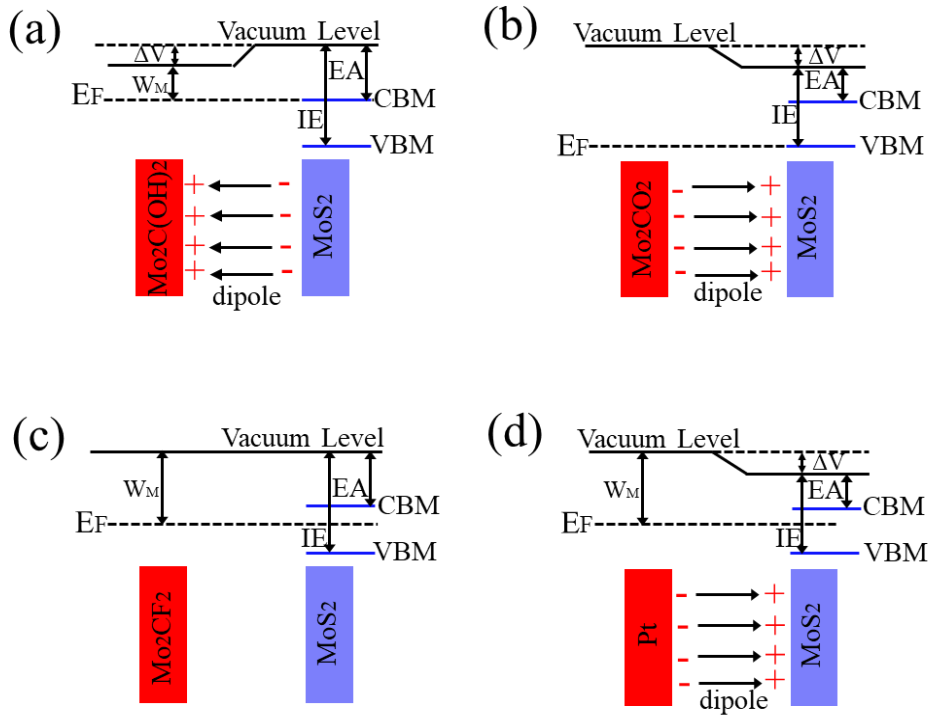

**FIG. 3.** Schematic illustration of band alignment for 1L MoS$_2$-Mo$_2$C(OH)$_2$ (a), 1L MoS$_2$-Mo$_2$CO$_2$ (b), 1L MoS$_2$-Mo$_2$CF$_2$ (c), 1L MoS$_2$-Pt (d).

The charge transfer take place at interface of Mo$_2$C(OH)$_2$-MoS$_2$, and Mo$_2$CO$_2$-MoS$_2$. The work function of Mo$_2$C(OH)$_2$ is lower than EA of MoS$_2$ while Mo$_2$CO$_2$ is larger than IE. In the first case (Figure 2a), electrons would be transferred from the 2D metal Mo$_2$C(OH)$_2$ to MoS$_2$, and then MoS$_2$ becomes negatively charged which is further confirmed by Bader charge [32] calculation that MoS$_2$ obtain 0.11 electrons. This creates a negative dipole (Formula 5) at the interface that shifts upwards the vacuum level as shown in Figure 3a. The electron continues transfer until the value of the CBM level of MoS$_2$ aligns with the Fermi level of the metal. An n-type vanishing SBH is obtained (see

Figure 3a and Figure S2a). In Figure 2b, on the contrary, electrons transferred from MoS$_2$ to the 2D metal Mo$_2$CO$_2$ with Bader charge result that MoS$_2$ lose 0.34 electrons, for Mo$_2$CO$_2$ work function is larger than IE. Similarly, this charge transfer also creates a dipole at the interface which is positive based on Formula 5 and reduces the vacuum level until equilibrium is reached. Therefore, a p-type vanishing SBH is obtained (see Figure 3b and Figure S2b).

Figure 2c shows charge redistribution at Mo$_2$CF$_2$-MoS$_2$ interface. The work function of Mo$_2$CF$_2$ is situated between EA and IE of MoS$_2$, therefore when MoS$_2$ adsorbed on Mo$_2$CF$_2$, there is nearly no charge transfer across the interface, no dipole formation, and the vacuum levels of the Mo$_2$CF$_2$ and MoS$_2$ remain aligned. However, we found that at Mo$_2$CF$_2$-MoS$_2$ interface, electrons localized in the interlayer region and deplete at both fluorine and sulfur atomic layer which is consistent with the electronic interlayer hybridization in 2L PtS$_2$ and few-layered black phosphorus called "covalent-like quasi-bonding" [33]. The electron aggregation area favors fluorine because fluorine is more electronegative than sulfur. The "covalent-like quasi-bonding" occurs under the following conditions: the interaction between the two materials forming heterojunction is weak, but strongly hybridized than simple vdW; and the electronegativity of the atomic layers at the interface is similar to each other. Mo$_2$CF$_2$-MoS$_2$ junction satisfies the above conditions and forms a covalent-like quasi-bonding which has no effect on interface coupling formation. The SBH in Mo$_2$CF$_2$-MoS$_2$ junction conform to Formula 1 which is called Schootky-Mott limit.

In conclusion, the symmetry of wave function on both sides of the interface determined that the push back effect can only occur in the 3D metal-MoS$_2$ heterojunction, since 2D metal Mo$_2$CT$_2$ and 2D MoS$_2$ have the similar ability to offer and accept electrons, the electrons spilled from the surface were evenly distributed on both sides of the interface in 2D metal-MoS$_2$ heterojunction. Charge transfer occurs when the metal work functions are extremely large and small for 2D metal. For different metal substrates, different charge redistribution mechanisms play a major role, resulting in the final interfacial ΔV.

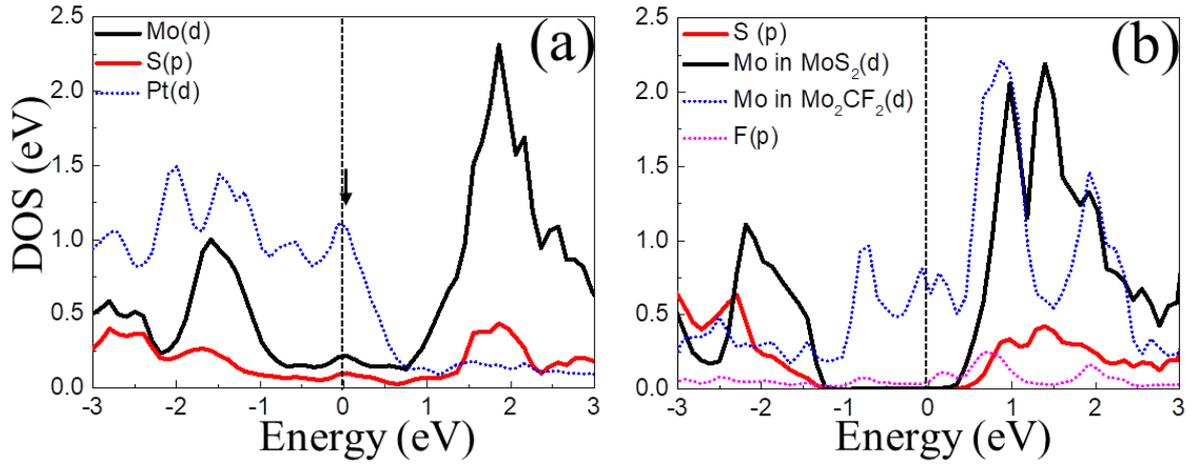

**FIG. 4.** Partial DOS of $MoS_2$-Pt (a), $MoS_2$-$Mo_2CF_2$ (b). The dashed arrow in panel (a) highlights the metal-induced states in $MoS_2$.

Besides charge redistribution, the metal induced gap states also play significant role in 3D metal-$MoS_2$ junctions in forming $\Delta V$ [34]. The partial density of states (DOS) of $MoS_2$ in Pt-$MoS_2$ and $Mo_2CF_2$-$MoS_2$ are addressed in Figure 4 to explore the electronic band offset induced by the interface. It can be seen that a distinct overlap between the Pt 5d, Mo 4d, and S 3p orbitals can be seen near the Fermi level, as indicated by the dashed arrow in Figure 4a, suggesting strong hybridization upon interface formation [21]. Additionally, metal-induced states (MIGS) appear in the partial DOS of both Mo and S atoms in $MoS_2$ layer, especially the Mo layer, when $MoS_2$ absorbed on Pt surface, in spite of the hybridization of Pt states with the $MoS_2$ states is not negligible. The MIGS also cause charge redistribution, which will induce a dipole $\Delta$ at the interface and finally screen the difference between SBH and $\Phi_B$. We note that $MoS_2$ layer in $Mo_2CF_2$-$MoS_2$ remains semiconducting with a band gap, and there are no metal induced gap states in $MoS_2$ layer which means no charge flow at M-S interface.

We have discussed the interfacial mechanism of charge redistribution at the M-S interface including push back effect, charge transfer, covalent-like quasi-bonding, and MIGS, which will eventually lead to the formation of interface couple $\Delta$, as shown in Formula 4. The interface couple $\Delta$ modifies SBH from the Schottky –Mott limit (Formula 1) to Formula 2; therefore we can presume SBH in metal -1L $MoS_2$ heterojunctions. For 2D metal substrates, when metal work function is smaller than the EA or is larger than IE of $MoS_2$ ($Mo_2C(OH)_2$ or $Mo_2CO_2$), the Fermi level of metal -1L $MoS_2$ heterojunctions are pinned at CBM or VBM of MoS$_2$ which means 0 n-type or p-type SBH; when the work function of the substrate is larger than the EA, but smaller than the IE, there is no charge transfer across the interface, no dipole formation, and the vacuum levels of the Mo$_2$CF$_2$ and MoS$_2$ remain aligned, which is called Schottky-Mott limit, where the SBH can be calculated by Formula 1. The 3D metal substrates and MoS$_2$ have a very complex interaction which leads to an unpredictable SBH. Both push back effect and MIGS play major role in forming SBH, which creates a positive dipole at the interface for the common used 3D metal substrates (like Ag, Au, Cu, Pt, the work function of which is located between EA and IE of 1L MoS$_2$).

Table II. SBHs of Metal−1L MoS$_2$ junctions read from projected band structures and calculated from formula 2 and formula 1.

|  | SBH from projected band structures (eV) | SBH from formula 2 (eV) | SBH from formula 1 (eV) |
| --- | --- | --- | --- |
| Mo$_2$C(OH)$_2$-MoS$_2$ | 0.01 | 0.24 | -2.20 |
| Mo$_2$CO$_2$-MoS$_2$ | -0.09 | 0.07 | -1.59 |
| Mo$_2$CF$_2$-MoS$_2$ | 0.38 | 0.47 | 0.49 |
| Pt-MoS$_2$ | 0.79 | 0.95 | 1.67 |
| Ag-MoS$_2$ | 0.14 | 0.26 | 0.4 |

The projected band structures of the MoS$_2$ supported on 2D metals Mo$_2$CT$_2$ and 3D metal Ag, Pt are shown in Figure S2 and S4, where SBHs are also labeled. From Table II, the SBHs calculated from projected band structures are close to the ones from formula 2 which considers the effects of Δ besides Schottky-Mott limit (formula 1). Δ has little effects on SBH in Mo$_2$CF$_2$-MoS$_2$, since there is neither MIGS nor charge transfer at M-S interface. Except in Mo$_2$CF$_2$-MoS$_2$, interface dipole Δ caused by interfacial charge distribution has a great effect on SBH in all other junctions.

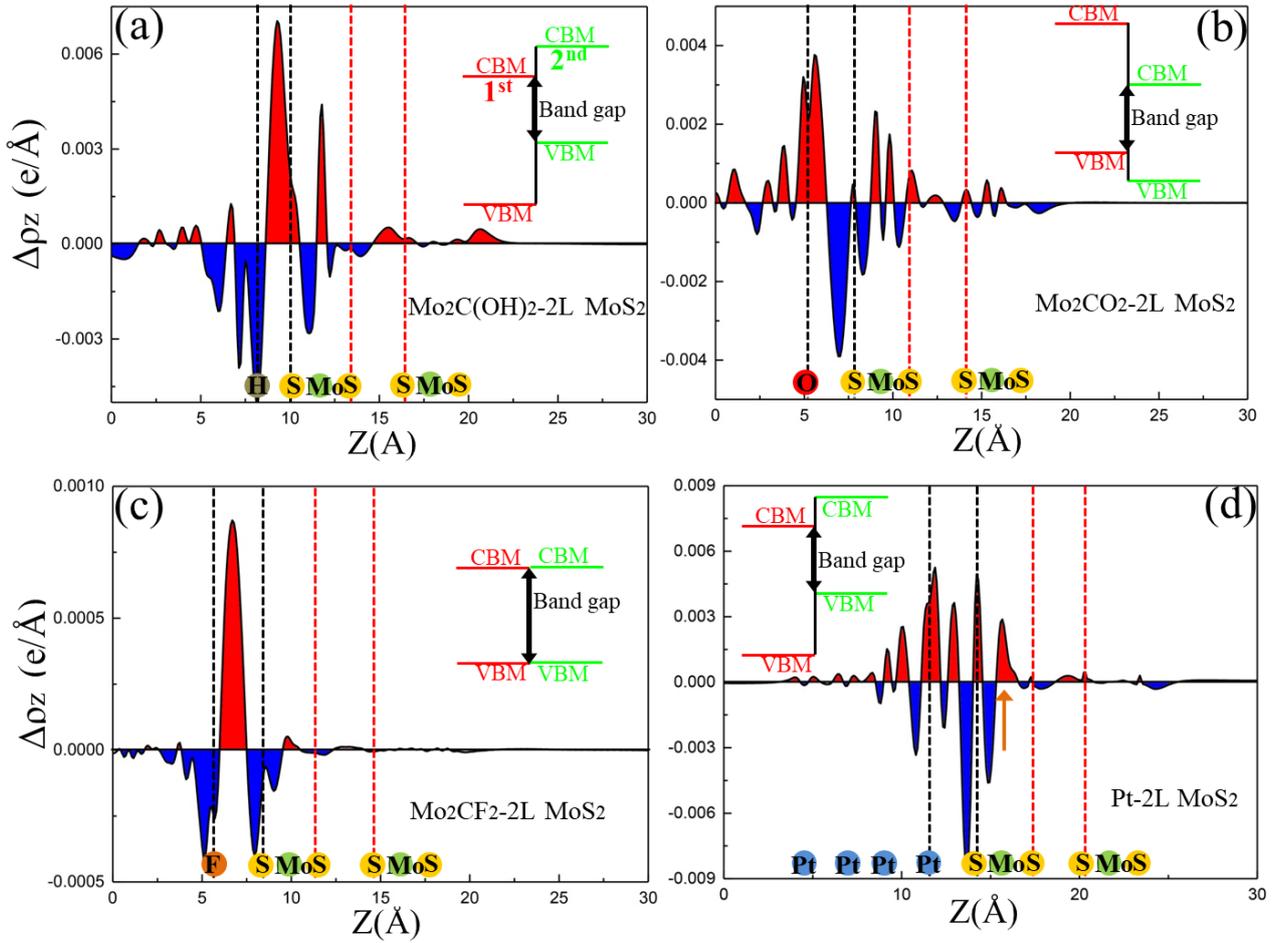

**FIG. 5. Charge distribution at Metal-2L MoS$_2$ interface.** (a−d) Plane-averaged electron density difference along the vertical z-direction to the Metal-2L MoS$_2$ interfaces. Red (blue) regions represent electron accumulation (depletion) regions. The interface between metal and MoS$_2$ is demonstrated within two black dotted lines, while interface between two MoS$_2$ layers is demonstrated within two red dotted lines. The insert images in (a-d) illustrate the type II band alignment between MoS$_2$ layers, where red (green) line indicates energy bands from the first (second) layer.

We have discussed the charge distribution at M-S interface, and found that various charge distributions will appear based on different metal substrates which ultimately affect the formation of SBH. Here, S-S interfaces are focused as shown in Figure 5 since the charge distribution at the S-S interface will give rise to band offset between the two layers of MoS$_2$. Before that, we have compare the charge distribution at M-S interface in metal-2L MoS$_2$ junction (Figure 5) and in metal-1L MoS$_2$ junction (Figure 2), and found that the charge distribution is almost identical between these two M-S interfaces.

The charge transfer condition between $Mo_2C(OH)_2$ (or $Mo_2CO_2$) and 2L $MoS_2$ is similar to that between $Mo_2C(OH)_2$ (or $Mo_2CO_2$) and 1L $MoS_2$ since the band alignments between them are same (the work function of $Mo_2C(OH)_2$ is smaller than EA of 2L $MoS_2$, while the work function of $Mo_2CO_2$ is larger than IE of 2L $MoS_2$). Acting as a continuation of the M-S interface, the charge transfer at S-S interfaces in these two junctions is same as that at M-S interfaces correspondingly, with the amount decreased considerably. In $Mo_2C(OH)_2$-2L $MoS_2$ shown in Figure 5a and 6a, electron accumulate at the $2^{nd}$ layer $MoS_2$ side, therefore the energy band of the $2^{nd}$ layer moves up to the vacuum level, leading to a large band offsets $\Delta_{CBM}$ between $1^{st}$ and $2^{nd}$ at CBM as shown in the inset image [type-II]. On the contrary, In $Mo_2CO_2$-2L $MoS_2$ shown in Figure 5b and 6b, electrons delete at the $2^{nd}$ layer $MoS_2$ side and an opposite band offsets $\Delta_{CBM}$ is formed. Therefore, a perfect type II homojunction is proved, since the CBM and VBM in $Mo_2C(OH)_2$-2L $MoS_2$ and $Mo_2CO_2$-2L $MoS_2$ junctions are sharply located at the two different layers of $MoS_2$, respectively. In $Mo_2CF_2$-2L $MoS_2$ junction, the amount of charge redistribution at M-S interface is very small, so the charge redistribution at the S-S interface, caused by which at M-S interface, is negligible. The overlapping energy bands of $1^{st}$ and $2^{nd}$ $MoS_2$ layers also prove this (see Figure S3c).

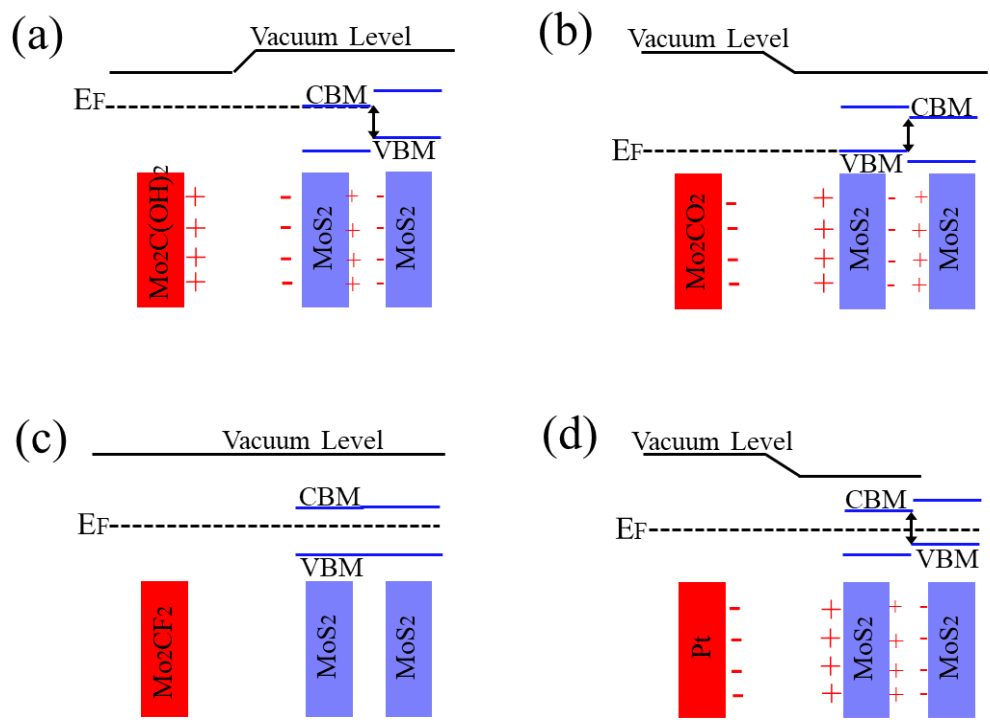

**FIG. 6.** Schematic illustration of band alignment for 2L $MoS_2$-$Mo_2C(OH)_2$ (a), 2L $MoS_2$-$Mo_2CO_2$ (b), 2L $MoS_2$-$Mo_2CF_2$ (c), 2L $MoS_2$-Pt (d).

For S-S interface in 3D metal Pt-2L MoS$_2$ junction, a different mechanism appears relative to 2D metal-2L MoS$_2$ junction, which cause an opposite charge distribution at S-S interface from that at M-S interface as shown in Figure 5d and 6d. As discussed, MIGS appears in 1$^{st}$ MoS$_2$, this also means partial metallization in 1$^{st}$ MoS$_2$. Metal screening effect occurs inside the 1$^{st}$ MoS$_2$ layer and hence it can screen the dipole at the metal-MoS$_2$ interface, which is indicated by the arrow in Figure 5d [depinning]. Originated from the screening effect inside 1$^{st}$ MoS$_2$ layer, at S-S interface, the electrons deplete at the left side and accumulate at the right side. Similarly, electrons also shift the energy band of 2$^{nd}$ layer MoS$_2$ to the vacuum level, which lead SBH turn from n-type to p-type, shown in Figure 6d.

## Conclusions

In conclusion, we have carried out first-principles calculations of the charge distribution at M-S and S-S interfaces of monolayer and bilayer MoS$_2$ supported on a series of metal electrodes. At M-S interfaces, push back effect plays a major role for M-S interface in 3D metal-MoS$_2$ junctions; charge transfer and covalent like bond appears in 2D metal-MoS$_2$ junctions based on band alignment of metal and MoS$_2$. At S-S interfaces, the charge distribution is same as that at M-S interfaces correspondingly in 2D metal-2L MoS$_2$; an opposite charge distribution at S-S interface from that at M-S interface is found in 3D metal-2L MoS$_2$. The charge distribution at M-S interface caused interface dipole ΔV and ultimately affected the SBH at the M-S interfaces; while the charge distribution at S-S interface gave rise to band offset between the two layers of MoS$_2$. In summary, our work offered a comprehensive study of interfacial interaction at M-S and S-S interface based on metal-1L/2L MoS$_2$, which stimulates the study of properties and applications in metal-multilayer semiconductor junctions.